\documentclass[final, journal]{IEEEtran}
\usepackage{amsfonts}
\usepackage{float}
\usepackage{amssymb}

\usepackage{algorithm}
\usepackage{algorithmic}
\usepackage{multirow}

\usepackage{amsthm}
\usepackage{CJK}
\usepackage{bookmark}
\usepackage{booktabs}
\usepackage{authblk}
\usepackage{setspace}

\usepackage{cite}

\ifCLASSINFOpdf
  \usepackage[pdftex]{graphicx}
   \graphicspath{{../pdf/}{../jpeg/}}
  \DeclareGraphicsExtensions{.pdf,.jpeg,.png}
\else
  \usepackage[dvips]{graphicx}
  \graphicspath{{../eps/}}
  \DeclareGraphicsExtensions{.eps}
\fi
\usepackage[cmex10]{amsmath}
\begin{document}
%
\title{A Low Complexity Approach of Combining Cooperative Diversity and Multiuser
Diversity in Multiuser Cooperative Networks}
%
%

\author{Tiejun~Lv,~\IEEEmembership{Senior Member,~IEEE,}~Zhang~Zhang,~and~Shaoshi~Yang,~\IEEEmembership{Member,~IEEE}

\thanks{Copyright (c) 2013 IEEE. Personal use of this material is permitted. However, permission to use this material for any other purposes must be obtained from the IEEE by sending a request to pubs-permissions@ieee.org.
This work is financially supported by the National Natural Science
Foundation of China (NSFC) (Grant No. 61271188).}
\thanks{T. Lv and Z. Zhang are with the School of Information
and Communication Engineering, Beijing University of Posts and Telecommunications,
Beijing 100876, China (e-mail: zhangzhang86@bupt.edu.cn; lvtiejun@bupt.edu.cn).}
\thanks{S. Yang is with the School of Electronics and Computer Science, University
of Southampton, SO17 1BJ Southampton, U.K., and also with
the School of Information and Communication Engineering, Beijing University
of Posts and Telecommunications, Beijing 100876, China (e-mail:
sy7g09@ecs.soton.ac.uk).}}

\markboth{To be published in IEEE Transactions on Signal Processing, 2013}%
 {Shell \MakeLowercase{\textit{et al.}}: Bare Demo of IEEEtran.cls
 for Journals}

\maketitle

\begin{abstract}
In this paper, we investigate the scheduling scheme to combine cooperative diversity (CD) and multiuser diversity (MUD) in multiuser cooperative networks under the time resource allocation (TRA) framework in which the whole transmission is divided into two phases: the broadcast phase and the relay phase. The broadcast phase is for direct transmission
whereas the relay phase is for relay transmission. Based on this TRA
framework, a user selection based low complexity relay protocol (US-LCRP)
is proposed to combine CD and MUD. In each time slot (TS) of the broadcast
phase, a ``best'' user is selected for transmission in order to
obtain MUD. In the relay phase, the relays forward the messages of
some specific users in a fixed order and then invoke the limited feedback
information to achieve CD. We demonstrate that the diversity-multiplexing
tradeoff (DMT) of the US-LCRP is superior to that of the existing
schemes, where more TSs are allocated for direct transmission in order
to jointly exploit CD and MUD. Our analytical and numerical results
show that the US-LCRP constitutes a more efficient resource utilization
approach than the existing schemes. Additionally, the US-LCRP can
be implemented with low complexity because only the direct links'
channel state information (CSI) is estimated during the whole transmission.\end{abstract}

\section{Introduction}

Diversity serves as one of the major solutions to combat channel impairment
caused by random fading in wireless environments \cite{1}. Recently,
cooperative communication has emerged as a promising technique of
achieving spatial diversity in a distributed fashion. A variety of
cooperation schemes such as opportunistic relaying and space-time
coded cooperation \cite{2,3,4,5} have been proposed to provide full
cooperative diversity (CD) in multi-relay networks. Among these schemes,
opportunistic relaying achieves full CD by selecting the ``best\textquotedblright{}
relay to support transmission. Moreover, it is outage-optimal under
an aggregate power constraint, and can be implemented with low complexity,
hence it attracts much attention.

On the other hand, it is well known that multiuser diversity (MUD)
constitutes an inherent resource of diversity in a multiuser network
\cite{6}. Since many users experience independent fading, the probability
that the ``best'' user has a ``strong'' channel is very high.
Therefore, by allowing only the user with the highest instantaneous
signal-to-noise ratio (SNR) to transmit, MUD can be obtained to improve
the outage probability and/or capacity performance.

In multiuser cooperative networks, it is potentially feasible to achieve
both CD and MUD, and there have been some studies focusing on the
combination of CD and MUD \cite{7,8,9,10,11,12,SR11,SR12,SR21,SR22,SR23,SR24}.
More specifically, \cite{7} and \cite{8} discussed this combination
in some specific cooperative networks from the capacity perspective,
other literature investigated the reliability performance of the combined
use of CD and MUD. The authors of \cite{9} established a multiuser
cooperative network model where each user transmits with the aid of
one exclusive relay, and analyzed the diversity order for both the
amplify-and-forward (AF) and the decode-and-forward (DF) protocols.
Furthermore, in \cite{10}, they extended the analysis of \cite{9}
to a more generalized multiuser network model in which each user has
multiple exclusive relays. However, the assumption of exclusive relay
might not be realistic although it brings convenience to theoretical
analysis. The authors of \cite{11} considered a more practical scenario
where all the users share all the relays, and proposed an optimal
``user-relay'' pair selection strategy to achieve CD and MUD simultaneously.
Nevertheless, global channel state information (CSI) is needed to
perform such ``user-relay'' pair selection. Namely, in an $N$-user
$M$-relay network, the CSI of all the $N\left(M+1\right)$ links
in the network is required for a single ``user-relay'' pair transmission\cite{12}.
This requirement makes the complexity of selection excessively high
for large $N$ and $M$. To reduce the complexity, the authors of
\cite{12} proposed a two-step selection scheme while still obtaining
both CD and MUD. To elaborate a little further, firstly, the ``best''
user with the highest direct-link channel quality is selected to transmit,
then a ``best'' relay is chosen to support the transmission. In
this way, only the CSI of the $N$ direct links and the $2M$ links
related to the relays are needed for the user selection and the relay
selection, respectively. The existing studies \cite{11,12} are based
on the time-resource allocation (TRA) framework that two time slots
(TSs) are allocated for each transmission request. In the first TS
the selected user broadcasts its information, and then in the second
TS the selected relay forwards its observation. However, considering
the two TSs as a whole, the framework is essentially the same as those
in the traditional non-cooperative systems. In this framework, using
two TSs together to serve one user causes a degradation of spectrum
efficiency.

Recently, a two-phase TRA framework (TP-TRA) is exploited to improve
the spectrum efficiency \cite{peng2007performance,key-2,key-4,my}.
In TP-TRA the whole transmission is divided into two phases: the broadcast
phase and the relay phase. Firstly, the users broadcast their messages
in the broadcast phase, and then the relays assist in transmission
in the relay phase. \cite{peng2007performance} showed that all the
users can achieve a diversity order of two by transmitting a network
coding (NC) combined packet within one relay TS in single DF-relay
aided systems. \cite{key-2,key-4} studied the cooperative schemes
in general networks with multiple users and multiple DF relays. More
specifically, \cite{key-2} proposed a Galois field NC based scheme
to achieve full CD, and \cite{key-4} developd a criterion for binary
field NC to guarantee full CD. In \cite{my}, the authors showed that
the diversity gain of NC based cooperation comes from selection, and
based on this revelation, they further proposed a user selection strategy
in the relay phase to achieve full CD for both AF and DF networks.
Attributing to the TP-TRA framework, these full-CD schemes can improve
the spectrum efficiency significantly. However, the problem of jointly
exploiting CD and MUD in multiuser multi-relay cooperative networks
has not been studied yet under the TP-TRA framework.

In this paper, we propose a user selection based low complexity relay
protocol (US-LCRP) which is capable of achieving both CD and MUD under
the TP-TRA framework. In each broadcast TS, the ``best'' user with
the strongest direct link broadcasts its data block. Then in the relay
phase, all the relays serve the transmission in a round-robin fashion.
In each relay TS, instead of selecting relay, the destination selects
a ``worst'' data block which most needs to be relayed according
to the quality record of each block. Afterwards, a single relay transmits
its observation of the selected data block, and the destination performs
data combining and quality record updating. To show the effectiveness
of the US-LCRP, the diversity-multiplexing tradeoff (DMT) performance
is analyzed in this paper.

The merit of the US-LCRP is twofold:
\begin{enumerate}
\item With the aid of good design, the US-LCRP achieves higher spectrum
efficiency while obtaining both CD and MUD. To be more specific, let
us suppose the broadcast phase and the relay phase last for $L$ and
$M$ TSs, respectively. Then, the US-LCRP provides better DMT performance
in the scenario of $L>M$, which indicates that the proposed protocol
achieves higher spectrum efficiency than the existing protocols \cite{11,12}
while maintaining the same reliability performance, or it attains
higher diversity gain than the existing protocols \cite{11,12} despite
providing the same data rate.
\item The US-LCRP requires the CSI of only the $N$ direct links for user
selection in each broadcast TS. The data block selection in the relay
phase is based on the quality record of the previously transmitted
data blocks. This quality record may be simply characterized as the
SNR of the corresponding signals received at the destination, and
can be estimated by using some SNR estimation algorithms. Hence the
US-LCRP imposes a significantly lower implementation complexity in
practice compared with existing protocols.
\end{enumerate}
The rest of this paper is organized as follows. The system model and
the proposed US-LCRP are described in Section II. The DMT performance
of the US-LCRP is analyzed in Section III, and simulation results
are provided in Section IV. Finally, conclusions are drawn in Section
V.

\section{System Model and The Proposed US-LCRP}

\subsection{System Model}

\begin{figure}[tbp]
\begin{centering}
\includegraphics[width=3.6in]{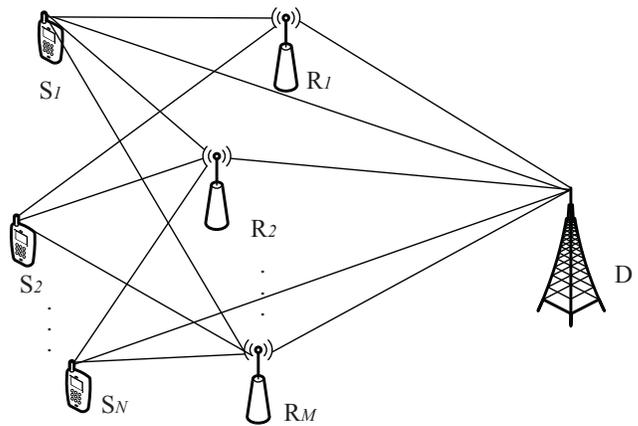}
\par\end{centering}
\caption{System Model.}
\label{fig:system_model}
\end{figure}

We consider an AF cooperative network with $N+M+1$ nodes, where $N$
users ($S_{n},1\leq n\leq N$ ) transmit individual information to
one destination ($D$) with the aid of $M$ relays ($R_{m},1\leq m\leq M$).
The system model is shown in Fig. \ref{fig:system_model}. All the nodes are assumed to
have single antenna and transmit with power $E_{s}$, and operate
in half-duplex mode. All the channels in the network are assumed to
be independent flat Rayleigh block fading channels with additive white
Gaussian noise (AWGN). We further assume that the variances of the
channel coefficients of the $S_{n}\rightarrow D$, $S_{n}\rightarrow R_{m}$,
and $R_{m}\rightarrow D$ links are $\gamma_{S_{n}D}$, $\gamma_{S_{n}R_{m}}$,
and $\gamma_{R_{m}D}$, respectively, while the average noise power
of each link in the network is $N_{0}$.

\subsection{The Proposed US-LCRP}

In \cite{11} and \cite{12}, the relays serve one data block immediately
after the data block's direct transmission has been finished. In this
TRA, the relays are dedicated to assist one data block in each relay
time slot and brings no benefits for the other data blocks. Differently,
we exploit the TRA as in \cite{peng2007performance,key-2,key-4,my}
where all the relays are shared by all the sources. To achieve this
effect, the relays do not participate in assisting the signal transmission
until all the direct transmissions have been finished. Therefore,
the whole transmission is divided into two phases: the broadcast phase
and the relay phase. First, the sources transmit data blocks in the
broadcast phase. Afterwards, the relays assist the transmissions in
the relay phase. Studies show that with the aid of a well-designed
protocol, it is attractive to achieve the effect of ``relay sharing\textquotedblright{},
which means that multiple sources are able to benefit from a single
relay TS. The design objective of US-LCRP is to achieve both CD and
MUD under the TP-TRA framework.

Different from \cite{11,12}, in the proposed US-LCRP, we exploit
the TRA framework as in \cite{peng2007performance,key-2,key-4,my}.
The whole transmission is divided into two phases: the broadcast phase
and the relay phase. First, the sources transmit data blocks in the
broadcast phase. Afterwards, the relays assist in transmissions in
the broadcast phase. Studies show that with the aid of a well-designed
protocol, it is attractive to achieve the effect of ``relay sharing''
which means that multiple sources are able to benefit from a single
relay TS. The design objective of US-LCRP is to achieve both CD and
MUD under the TP-TRA framework.

We assume that the broadcast phase occupies $L$ TSs. In each TS of
the broadcast phase, the ``best'' user whose link towards the destination
exhibits the highest SNR is selected as a candidate for transmission.
Then in the relay phase, the relays assist in transmissions one by
one, thus the relay phase lasts for $M$ TSs. In each relay TS, a
single relay aids the transmission of the ``worst'' data block which
has the lowest quality record at $D$. Fig. \ref{Fig2}(a) illustrates TP-TRA,
and its details are presented as follows.

\subsubsection{Broadcast Phase}

In the broadcast phase, a greedy scheduler is employed to obtain MUD.
In the $l$th broadcast TS, the scheduler chooses the ``best'' user
$S_{i_{l}}$ whose link towards the destination has the highest SNR.
Then, $i_{l}$ can be expressed as
\begin{equation}
i_{l}=arg\underset{n=1,\ldots,N}{\max}\rho_{n}^{\left(l,\mathrm{BP}\right)},\label{eq:UserSelection}
\end{equation}
where ``BP'' is the abbreviation of ``broadcast phase'', $\rho_{n}^{\left(l,\mathrm{BP}\right)}=\frac{E_{s}\left|h_{S_{n}D}^{\left(l,\mathrm{BP}\right)}\right|^{2}}{N_{0}}$
represents the instantaneous SNR of the link $S_{n}\rightarrow D$
in the $l$th broadcast TS, and $h_{S_{n}D}^{\left(l,\mathrm{BP}\right)}$
denotes the channel coefficient of this link in the $l$th broadcast
TS.

Due to the broadcast nature of wireless environment, all the relays
and the destination can receive $S_{i_{l}}$'s signal. The received
signal at $D$ and $R_{m}$ are
\begin{align*}
y_{S_{i_{l}}D} & =h_{S_{i_{l}}D}^{\left(l,\mathrm{BP}\right)}x_{l}+n_{S_{i_{l}}D}^{\left(l,\mathrm{BP}\right)},\\
y_{S_{i_{l}}R_{m}} & =h_{S_{i_{l}}R_{m}}^{\left(l,\mathrm{BP}\right)}x_{l}+n_{S_{i_{l}}R_{m}}^{\left(l,\mathrm{BP}\right)},
\end{align*}
respectively, where $h_{S_{i_{l}}R_{m}}^{\left(l,\mathrm{BP}\right)}$
is the channel coefficient of the link $S_{i_{l}}\rightarrow R_{m}$
in the $l$th broadcast TS, $x_{l}$ is the transmitted data block
of $S{}_{i_{l}}$ in the $l$th broadcast TS, $n_{S_{i_{l}}D}^{\left(l,\mathrm{BP}\right)}$
and $n_{S_{i_{l}}R_{m}}^{\left(l,\mathrm{BP}\right)}$ are the AWGN
at $D$ and $R_{m}$, respectively.

\subsubsection{Relay Phase}
\begin{figure}[tbp]
\begin{centering}
\includegraphics[width=3.5in]{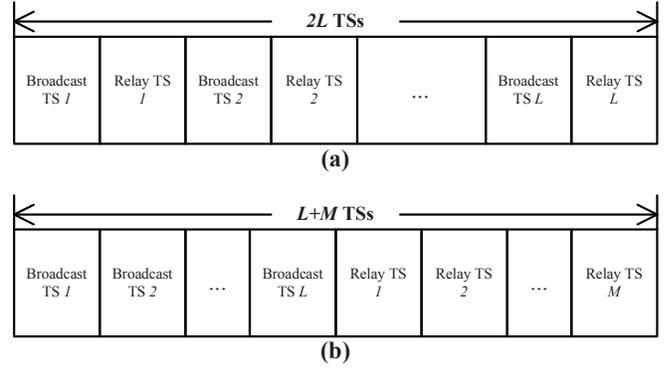}
\par\end{centering}
\caption{TRA framework of (a) the existing schemes \cite{11,12} and (b) the
proposed US-LCRP.}
\label{Fig2}
\end{figure}
The relay phase lasts for $M$ TSs, during which all the relays participate
in the transmission one by one, i.e., in a round-robin fashion. In
the first relay TS, $R_{1}$ transmits, and then in the second relay
TS, $R_{2}$ transmits. This procedure goes on until all the relays
have assisted the transmission. In addition, the US-LCRP employs data
block selection to facilitate the transmission. Briefly speaking,
a single relay assists the transmission of the ``worst'' data block
in each relay TS. After relays' transmission, the destination performs
data combining and then updates the quality record of the selected
data block in order to prepare for the next relay TS. Since selective
combining (SC) is capable of providing diversity order with rather
low complexity, we focus on SC in this paper. It should be noted that
other combining schemes such as maximum ratio combining (MRC) \cite{CC}
and equal gain combining (EGC) can be readily introduced into the
US-LCRP in the same way. The details of relay phase operation are
described as follows.

We denote $\rho_{l}^{\left(m\right)}$ as the SNR of the received
signals at $D$ related to $x_{l}$ after combining and before the
$m$th relay TS (i.e., the SNR of the combined signals from all the
links over which $x_{l}$ has been transmitted to $D$ before the
$m$th relay TS), $\mathrm{\phi}_{m}$ as the set of $\rho_{l}^{\left(m\right)}$,
where $1\leq l\leq L$. Suppose $h_{S_{n}R_{m}}^{\left(m,\mathrm{RP}\right)}$,
$h_{S_{n}D}^{\left(m,\mathrm{RP}\right)}$, and $h_{R_{m}D}^{\left(m,\mathrm{RP}\right)}$
are the channel fading coefficients of the links $S_{n}\rightarrow R_{m},$
$S_{n}\rightarrow D,$ and $R_{m}\rightarrow D$ in the $m$th relay
TS, respectively, where ``RP'' is the abbreviation of ``relay phase''.
For the ease of exposition, the details of the calculation and the
updating of $\rho_{l}^{\left(m\right)}$ will be explained later.

In the $m$th relay TS, $D$ first selects the ``worst'' data block
and broadcasts its index $\theta_{m}$,%
\footnote{This limited feedback information may be represented with $\left\lceil \log_{2}L\right\rceil $
bits.%
} where
\begin{align}
\theta_{m} & =\arg\underset{l=1,\ldots,L}{\min}\rho_{l}^{\left(m\right)},1\leq m\leq M.\label{eq:UserSelectionRelayPhase}
\end{align}

Then $R_{m}$ amplifies its observation of the data block $x_{\theta_{m}}$
and forwards it to $D$. The destination receives the relayed signal,
which is written as
\begin{equation}
y_{R_{m}D}=h_{R_{m}D}^{\left(m,\mathrm{RP}\right)}\widetilde{x}_{\theta_{m}}+n_{R_{m}D}^{\left(m,\mathrm{RP}\right)},
\end{equation}
where $n_{R_{m}D}^{\left(m,\mathrm{RP}\right)}$ is the AWGN with
zero mean and variance $N_{0}$, $\widetilde{x}_{\theta_{m}}$ is
the version of amplified signal $x_{\theta_{m}}$ at $R_{m}$ and
is expressed as
\begin{align}
\widetilde{x}_{\theta_{m}}= & \frac{h_{S_{i{}_{\theta_{m}}}R_{m}}^{\left(\theta_{m},\mathrm{BP}\right)}}{\sqrt{\left|h_{S_{i{}_{\theta_{m}}}R_{m}}^{\left(\theta_{m},\mathrm{BP}\right)}\right|^{2}+N_{0}}}x_{\theta_{m}}+\frac{n_{S_{i{}_{\theta_{m}}}R_{m}}^{\left(\theta_{m},\mathrm{BP}\right)}}{\sqrt{\left|h_{S_{i{}_{\theta_{m}}}R_{m}}^{\left(\theta_{m},\mathrm{BP}\right)}\right|^{2}+N_{0}}}.\label{eq:RelayedSignal}
\end{align}

After $R_{m}$ finishes the transmission, $D$ employs SC to perform
data combining. Finally, $D$ updates $\rho_{\theta_{m}}^{\left(m\right)}$
to $\rho_{\theta_{m}}^{\left(m+1\right)}$and reconstructs a new SNR
set $\phi_{m+1}$ to record all the data blocks' quality of the next
relay TS, and the current $m$th TSs ends.

The relay phase continues until all the relays finish assisting the
transmission and thereby lasts for $M$ TSs.

Below we will elaborate on the calculation and updating of $\rho_{l}^{\left(m\right)}$.
Note that the initial value of $\rho_{l}^{(m)}$, namely $\rho_{l}^{\left(1\right)}$,
is equal to the SNR of the received signals at $D$ from $S_{i_{l}}$
after the broadcast phase, and it is computed as
\begin{align}
\rho_{l}^{\left(1\right)} & =\rho_{i_{l}}^{(l,BP)}=\frac{E_{s}\left|h_{S_{i_{l}}D}^{\left(l,\mathrm{BP}\right)}\right|^{2}}{N_{0}}.\label{eq:SNRini}
\end{align}
In the $m$th relay TS, relying on (\ref{eq:RelayedSignal}), the
SNR of the received signal at $D$ (i.e., the SNR of $y_{R_{m}D}$)
is formulated as
\begin{align}
\rho^{\left(m,\mathrm{RP}\right)}= & \frac{E_{s}\left|h_{S_{i{}_{\theta_{m}}}R_{m}}^{\left(\theta_{m},\mathrm{BP}\right)}h_{R_{m}D}^{\left(m,\mathrm{RP}\right)}\right|^{2}}{N_{0}\left(\left|h_{S_{i{}_{\theta_{m}}}R_{m}}^{\left(\theta_{m},\mathrm{BP}\right)}\right|^{2}+\left|h_{R_{m}D}^{\left(m,\mathrm{RP}\right)}\right|^{2}+N_{0}\right)}\label{eq:SNRrelayedSignal}\\
= & \frac{\rho_{S_{i{}_{\theta_{m}}}R_{m}}\rho_{R_{m}D}}{\rho_{S_{i{}_{\theta_{m}}}R_{m}}+\rho_{R_{m}D}+1},\label{eq:Rou_Double}
\end{align}
where $\rho_{S_{i{}_{\theta_{m}}}R_{m}}$ is the SNR of the link $S_{i{}_{\theta_{m}}}\rightarrow R_{m}$
in the $\theta{}_{m}$th broadcast TS and can be expressed as $\rho_{S_{i{}_{\theta_{m}}}R_{m}}=\frac{E_{s}\left|h_{S_{i{}_{\theta_{m}}}R_{m}}^{\left(\theta_{m},\mathrm{BP}\right)}\right|^{2}}{N_{0}}$,
while $\rho_{R_{m}D}$ is the SNR of the link $R_{m}\rightarrow D$
in the $m$th relay TS and can be expressed as $\rho_{R_{m}D}=\frac{E_{s}\left|h_{R_{m}D}^{\left(m,\mathrm{RP}\right)}\right|^{2}}{N_{0}}.$

After data combining, $\rho_{l}^{\left(m\right)}$ is updated to $\rho_{l}^{\left(m+1\right)}$
. Additionally, $\rho_{\theta_{m}}^{\left(m+1\right)}$ denotes the
SNR of the combined signal at the output of the selection combiner,
thus it is written as
\begin{align}
\rho_{\theta_{m}}^{\left(m+1\right)}= & \max\left(\rho_{\theta_{m}}^{\left(m\right)},\rho^{\left(m,\mathrm{RP}\right)}\right).\label{eq:SNRupdating}
\end{align}
Since only the signal of the selected data block is relayed, the unselected
data blocks' quality records remain unchanged. Thus the elements of
the updated SNR set $\phi_{m+1}$ are given by
\begin{align}
\rho_{l}^{\left(m+1\right)}= & \begin{cases}
\rho_{l}^{\left(m\right)}, & l\neq\theta_{m}\\
\max\left(\rho_{l}^{\left(m\right)},\rho^{\left(m,\mathrm{RP}\right)}\right), & l=\theta_{m}
\end{cases}.\label{eq:SNRsetupdating}
\end{align}

\subsubsection{Discussions}

Let us compare the proposed US-LCRP and the existing protocols in
\cite{11,12}. The US-LCRP enjoys a lower implementation complexity.
In \cite{11}, the CSI of all the $N\left(M+1\right)$ links in the
entire network has to be estimated for selecting the optimal ``user-relay''
pair selection. In \cite{12}, the CSI of only the $N$ direct links
and of the $2M$ relaying links has to be estimated for selecting
the ``proper'' user and the ``proper'' relay. In contrast, the
US-LCRP requires to estimate the CSI of only the $N$ direct links
in the broadcast phase while in the relay phase it does not need to
estimate any CSI. In each relay TS of the US-LCRP, the destination
$D$ only has to evaluate the received SNR and update the users' quality
records. It should be noted that in each relay TS of the scheme proposed
in \cite{12}, $D$ also has to estimate the received SNR of the relayed
signal as in the proposed US-LCRP scheme, because it also has to combine
the original signal and the relayed signal. Besides, only the quality
of the selected ``data block\textquotedblright{} is changed and updated
in each relay TS of the proposed US-LCRP scheme, and the updating
can be directly accomplished after the data combining, as shown in
(\ref{eq:SNRsetupdating}). As a result, the calculation and updating
of each data block's quality do not incur other overhead when compared
with the scheme of \cite{12}.

Moreover, Fig. \ref{Fig2} illustrates the TRA of the US-LCRP and
its counterparts in \cite{11,12}. We can observe from Fig. \ref{Fig2}
that the US-LCRP occupies $L+M$ TSs whereas both schemes of \cite{11}
and \cite{12} occupy $2L$ TSs. It implies that if $L$ is set a
larger value than $M$, the US-LCRP has the potential to achieve higher
spectral efficiency than the schemes of \cite{11,12}, because the
US-LCRP requires less time resources. This conclusion will be demonstrated
by both the analytical results of Section III and the simulation results
of Section IV, as detailed subsequently.

Let us now discuss the issue of feedback latency. In each TS of
the proposed US-LCRP, the destination (i.e. BS) has to notify the
selected users/relays with a limited amount of feedback information.
It should be pointed out that such kind of feedback technique is widely
exploited in wireless communications, and related works include, for
example, the benchmark schemes of \cite{11} and \cite{12} considered
in this paper. In the scheme of \cite{11}, the destination performs
joint ``user-relay'' pair selection among all the $NM$ candidates
of the ``user-relay'' pairs. Therefore, the destination has to broadcast
$\left\lceil \log NM\right\rceil $ bits of feedback information to
reveal the selection result, where $\left\lceil .\right\rceil$ denotes the ceiling operation. In the scheme of \cite{12}, the destination
performs ``optimal'' user selection among all the $N$ candidates of
users and ``optimal'' relay selection among all the $M$ candidates
of relays in each broadcast time slot and each relay time slot, respectively.
$\left\lceil \log N\right\rceil$ and $\left\lceil \log M\right\rceil$
bits feedback information is thus used for notifying the selected users
and the selected relays, respectively. By contrast, in the proposed
US-LCRP requires smaller amount of feedback information compared with
its counterpart schemes in \cite{11} and \cite{12}, when $L\geq\max\left\{ M,3\right\} $.
Furthermore, feeding back a small amount of information is not difficult
in practical systems such as LTE, where the feedback information is
transmitted with much lower data rate and protected with much stronger
channel codes to ensure a much lower error rate compared with the
data since it is important and of small size. As a result, the feedback
information could be regarded as approximately perfect. What\textquoteright{}s
more important, the latency brought by feedback is typically acceptable
to satisfy the QoS requirement in current wireless communication systems.
For instance, in LTE the ACK/NACK is sent four subframes later than
the data block\textquoteright{}s transmission, and the retransmission
will happen four subframes later after the NACK is sent. In current
wireless systems, the time granularity of scheduling is tiny enough
and typically this latency does not affect user experience. In addition,
in this way the resources could be used efficiently. The uplink and
downlink usually use different frequency bands, and after the uplink/downlink
feedback information has been sent, the base station/user equipment
can use the downlink/uplink channel to transmit other information
to the user equipment/base station.

\section{DMT Analysis}

Both the reliability performance and the spectral efficiency are considered
to verify the superiority of the US-LCRP. As is well known, the fundamental
and comprehensive performance metric to simultaneously characterize
the reliability and capacity performance is DMT \cite{DMT02}. DMT
depicts the reliability with diversity gain and the capacity with
multiplexing gain. It shows the achievable diversity gain of a given
protocol under a certain multiplexing gain. In this section, the DMT
performance of the US-LCRP is analyzed. For convenience of exposition,
let us start with the definitions that will be used in our analysis.

\subsection{Definitions}

We define $\mathit{\mathsf{U}}=\left\{ u_{1},\ldots,u_{k},\ldots,u_{L+M}\right\} $
$\left(1\leq k\leq L+M\right)$ as the set of SNRs of the received
signals at $D$ in the broadcast phase and the relay phase. Considering
(\ref{eq:SNRini}), (\ref{eq:Rou_Double}) and (\ref{eq:SNRupdating}),
we have $\mathit{\mathsf{U}}=\left\{ \rho_{i{}_{l}}^{\left(l,\mathrm{BP}\right)}|l=1,2,\ldots,L\right\} \cup\left\{ \rho^{\left(m,\mathrm{RP}\right)}|m=1,2,\ldots,M\right\} $,
thus $u_{k}$ is defined as
\begin{equation}
u_{k}=\begin{cases}
\rho_{i{}_{k}}^{\left(k,\mathrm{BP}\right)}, & 1\leq k\leq L\\
\rho^{\left(k-L,\mathrm{RP}\right)}, & L<k\leq L+M
\end{cases}.\label{eq:DefUk}
\end{equation}
We assume the elements of the set $\mathit{\mathsf{U}}$ are ordered
as $u_{1}^{'}\le u_{2}^{'}\le\ldots\le u_{L+M}^{'}$.

Define $\mathsf{V}=\left\{ v_{l}|l=1,2,\ldots,L\right\} $, where
$v_{l}$ represents the SNR of the combined signals of $x_{l}$ at
$D$ after the whole transmission. Apparently,
\begin{equation}
v_{l}=\rho_{l}^{\left(M+1\right)}.\label{eq:DefVk}
\end{equation}
We also assume the elements of the set $\mathsf{V}$ are ordered as
$v_{1}^{'}\le v_{2}^{'}\le\ldots\le v_{L}^{'}$.

The average transmitted SNR of the network is defined as the effective
signal to noise power ratio, namely
\begin{align}
\rho & =\frac{LE_{s}}{\left(L+M\right)N_{0}},\label{eq:DefSNR}
\end{align}
where $E_{s}$ is the transmission power at each node.

A system outage event occurs when $D$ does not correctly decode all
the blocks after the whole transmission. Let us define $I_{l}$ as
the maximum average mutual information between $x_{l}$ and the corresponding
received signal at $D$. For a given end-to-end data rate of $\mathcal{R\,}bit/s/Hz$,
$S_{i_{l}}$ suffers an outage if $I_{l}=\frac{L}{L+M}log\left(1+v_{l}\right)<\mathcal{R}$.
Thus the system outage takes place if the condition $\min\left\{ I_{1},I_{2},\ldots,I_{L}\right\} \geq\mathcal{R}$
is not satisfied.

The multiplexing gain is defined as \cite{DMT02}
\begin{gather}
r=\lim\limits _{\rho\to\infty}\frac{\mathcal{R}\left(\rho\right)}{\log\rho},\label{eq:DefDiv}
\end{gather}
where $\mathcal{R}\left(\rho\right)$ is the end-to-end transmission
data rate characterized as a function of the average SNR $\rho$.

The diversity gain is defined as \cite{DMT02}
\begin{gather}
d=-\lim\limits _{\rho\to\infty}\frac{\log P_{out}\left(\mathcal{R}\left(\rho\right)\right)}{\log\rho},\label{eq:DefMul}
\end{gather}
where $P_{out}\left(\mathcal{R}\left(\rho\right)\right)$ is the average
system outage probability for the given $\rho$ and data rate $\mathcal{R}\left(\rho\right)$.
This definition is also written as $P_{out}\left(\mathcal{R}\left(\rho\right)\right)\doteq\rho^{-d}$
in the exponential equality notation as used in \cite{DMT02}.

\subsection{DMT performance}

Next, we proceed to analyze the DMT performance of our US-LCRP with
the above definitions. Obviously, we need to evaluate the system outage
probability with regard to the average SNR and the required end-to-end
data rate. Relying on the definition of the system outage event, the
system outage probability is expressed as
\begin{align}
P_{out}\left(\mathcal{R}\right) & =1-\Pr\left(\min\left\{ I_{1},I_{2},\ldots,I_{L}\right\} \geq\mathcal{R}\right)\nonumber \\
 & =1-\Pr\left(\min\left\{ v_{1},v_{2},\ldots,v_{L}\right\} \geq2^{\frac{L+M}{L}\mathcal{R}}-1\right)\nonumber \\
 & =1-\Pr\left(v_{1}^{'}\geq2^{\frac{L+M}{L}\mathcal{R}}-1\right)\nonumber \\
 & =\mathrm{F}_{v_{1}^{'}}\left(\beta\right),\label{eq:DefPouts}
\end{align}
where $\mathrm{F}_{v_{1}^{'}}\left(.\right)$ is the cumulative distribution
function (CDF) of $v_{1}^{'}$, $\beta\triangleq2^{\frac{L+M}{L}\mathcal{R}}-1$.

To begin with, let us study the equivalent SNR of the combined signal received
at $D$ for each data block after the whole transmission is finished. In total there are
$L+M$ data blocks transmitted during the whole transmission. Since
one data block is discarded while making data combining in each relay
TS, $L$ data blocks are retained for the final decision. First, we
prove that the retained data blocks have better quality than the discarded
data blocks. This conclusion is summarized in the following Lemma 1.
\newtheorem*{thm}{Lemma 1}
\begin{thm}
The SNRs of the retained data blocks are higher than that of the discarded
data blocks, i.e.,
\begin{equation}
v_{l}^{'}=u_{l+M}^{'},\label{eq:UVrelation}
\end{equation}
where $v_{l}^{'}$ is the $l$th smallest SNR of the retained data blocks and $u_{l+M}^{'}$ is the $l+M$th smallest SNR of all the data blocks as defined in
Section III-A previously.
\end{thm}
\begin{IEEEproof}
Please see Appendix A.
\end{IEEEproof}
With Lemma 1, (\ref{eq:DefPouts}) can be rewritten as
\begin{equation}
P_{out}\left(\mathcal{R}\right)=\mathrm{F}_{u_{M+1}^{'}}\left(\beta\right),\label{eq:CalPouts}
\end{equation}
where $\mathrm{F}_{u_{M+1}^{'}}\left(.\right)$ is the CDF of $u_{M+1}^{'}$,
and can be formulated as
\begin{align}
 & \mathrm{F}_{u_{M+1}^{'}}\left(\beta\right)\nonumber \\
= & \Pr\left(\mbox{at least \ensuremath{M+1\,}entries of \ensuremath{\mathit{\mathsf{U}}\,}are less than/equal to \ensuremath{\beta}}\right)\nonumber \\
= & \sum\limits _{j=M+1}^{L+M}\Pr\left(\mbox{\ensuremath{j\,}entries of \ensuremath{\mathit{\mathsf{U}}\,}are less than/equal to \ensuremath{\beta}}\right).\label{eq:lemma2CDF_lm}
\end{align}

To calculate (\ref{eq:lemma2CDF_lm}), we try to obtain the expression
of $\mathrm{F}_{u_{k}}\left(\beta\right)$. For $1\leq k\leq L$,
$\mathrm{F}_{u_{k}}\left(\beta\right)$ represents the CDF of the
SNR of the strongest direct link in a single broadcast TS. Thus it
is expressed as
\begin{align}
\mathrm{F}_{u{}_{k}}\left(\beta\right) & =\prod\limits _{n=1}^{N}\Pr\left(\rho_{n}^{\left(k,\mathrm{BP}\right)}<\beta\right)\label{eq:SymCDF1}\\
 & =\prod\limits _{n=1}^{N}\left(1-\exp\left(-\lambda_{S_{n}D}\beta\right)\right),
\end{align}
where $\lambda_{S_{n}D}=\frac{N_{0}}{E_{s}\gamma_{S_{n}D}}$. For
$L<k\leq L+M$, $\mathrm{F}_{u_{k}}\left(\beta\right)$ is the CDF
of the SNR of the relayed signal in the $\left(k-L\right)$th relay
TS. However, in general networks where the distributions of the SNRs
of all the links are characterized by different parameters, and hence
the distribution of $u_{k}$ is varying corresponding to the specific
values of $\theta_{k-L}$. As a result, $\mathrm{F}_{u_{k}}\left(\beta\right)$
is written as
\begin{align}
\mathrm{F}_{u{}_{k}}\left(\beta\right) & =\sum\limits _{l=1}^{L}\Pr\left(\theta_{k-L}=l\right)\mathrm{F}_{u{}_{k}}\left(\beta|\theta_{k-L}=l\right),\label{eq:SymCDF2}
\end{align}
where
\begin{align}
 & \mathrm{F}_{u_{k}}\left(\beta|\theta_{k-L}=l\right)\nonumber \\
= & 1-\exp\left(-\lambda_{S_{i_{\theta_{k-L}}}R_{k-L}}x-\lambda_{R_{k-L}D}x\right)\nonumber \\
 & \times2\sqrt{\lambda_{S_{i_{\theta_{k-L}}}R_{k-L}}\lambda_{R_{k-L}D}x\left(x+1\right)}\nonumber \\
 & \times K_{1}\left(2\sqrt{\lambda_{S_{i_{\theta_{k-L}}}R_{k-L}}\lambda_{R_{k-L}D}x\left(x+1\right)}\right),\label{eq:SymCDF2_x}
\end{align}
with $\lambda_{S_{n}R_{m}}$, $\lambda_{R_{m}D}$ being $\frac{N_{0}}{E_{s}\gamma_{S_{n}R_{m}}}$,
$\frac{N_{0}}{E_{s}\gamma_{R_{m}D}}$, respectively, and $K_{1}\left(.\right)$
being the first order modified Bessel function of the second kind.

With the expression of the system outage probability, we can derive
the DMT of the proposed US-LCRP. \newtheorem*{thm1}{Theorem 1}
\begin{thm1}
In an $N$-user $M$-relay network with multiplexing gain $r$, the
proposed US-LCRP protocol with $L$ broadcast phase TSs achieves the
DMT of
\begin{equation}
d=\left(M+N\right)\left(1-\frac{L+M}{L}r\right)^{+},\label{eq:DMTresu}
\end{equation}
where $\left(x\right)^{+}$ represents $\max\left\{ x,0\right\} $. \end{thm1}
\begin{IEEEproof}
Please see Appendix B.
\end{IEEEproof}

\subsection{Discussions}

\begin{figure}[tbp]
\begin{centering}
\textsf{\includegraphics[width=3.5in]{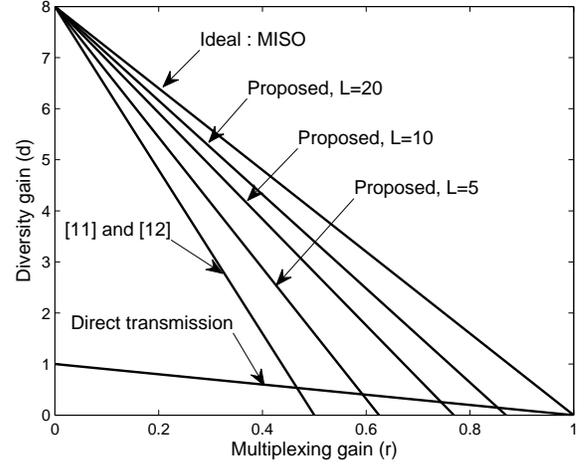}}
\par\end{centering}
\caption{An example of DMT performance comparison across the proposed US-LCRP,
the existing protocols of {[}11{]}, {[}12{]} and the ideal MISO scenario.
There are five sources and three relays in the network (i.e., $N=5$,
$M=3$) considered.}
\label{fig:DMTcurve}
\end{figure}

It is not difficult to prove that the existing protocols combining
CD and MUD \cite{11,12} achieve the DMT of $\left(M+N\right)\left(1-2r\right)^{+}$.
Clearly, the proposed US-LCRP provides a diversity order of $M+N$,
similar to the existing protocols. However, in the network where $L>M$,
the US-LCRP achieves better DMT performance. In other words, the US-LCRP
achieves higher spectral efficiency than the existing protocols while
maintaining the same level of reliability, or it provides higher diversity
gain than the existing protocols with the same data rate. Moreover,
as is well known, the ideal DMT of an $\left(M+N\right)$ -input one-output
network (an instance of the multiple-input single-output (MISO) network)
is $\left(M+N\right)\left(1-r\right)^{+}$. However, it is practically
infeasible to achieve such an ideal DMT in a cooperative network due
to the potential excessively high complexity. It is demonstrated that
as $L$ increases, the DMT of our US-LCRP approaches the ideal case.%
\footnote{(\ref{eq:DMTresu}) implies that as the value of $L$ goes to infinity,
the achievable DMT performance of the US-LCRP approaches that of the
ideal MISO scenario. In practical communication systems, the required
buffer size in each relay linearly grows upon increasing the value
of $L$, because each relay has to store all the $L$ data blocks.
Therefore, the value of $L$ should be set as large as possible subject
to the limitation of the buffer size of the relays.%
} As an example, we illustrate the DMT curves for various communication
strategies in a five-source three-relay network in Fig. \ref{fig:DMTcurve}.
When the multiplexing gain is $0.4$, the existing protocols combining
CD and MUD \cite{11,12} achieve a diversity gain of $1.6$, while
the US-LCRP achieves a diversity gain of $2.88$, for $L=5$. It demonstrates
that the US-LCRP achieves higher diversity gain than the existing
protocols while providing the same data rate. Moreover, the existing
protocols of \cite{11,12} achieve a diversity gain of $3.2$ with
multiplexing gain $r=0.3$, while our US-LCRP achieves a diversity
gain of $3.84$ and $4.32$ with multiplexing gain $r=0.4$ for $L=10$
and $L=20$, respectively. This result implies that the US-LCRP is
capable of offering higher data rate (i.e., larger multiplexing gain)
with higher reliability (i.e. larger diversity gain) if $L$ is set
to be sufficiently large.

\section{Simulation Results}

In this section, simulation results are presented to demonstrate the
performance of the proposed US-LCRP in terms of the average system
outage probability with different values of multiplexing gain. These
results corroborate the validity of the proposed protocol and consolidate
our DMT analysis presented in Section III. Here $E_{b}/N_{0}$ represents
the ratio of the average transmit power to the noise power of the
network. The simulations are performed over Rayleigh block fading
channels with AWGNs. The network is generated in a two-dimensional
plane where $D$ is located at the coordinate of $(1,1)$, and the
other nodes are uniformly distributed in the first quadrant of the
$1\times1$ square as \cite{12}. The path loss exponent is set to
$2$.

\begin{figure}[tbp]
\begin{centering}
\textsf{\includegraphics[width=3.5in]{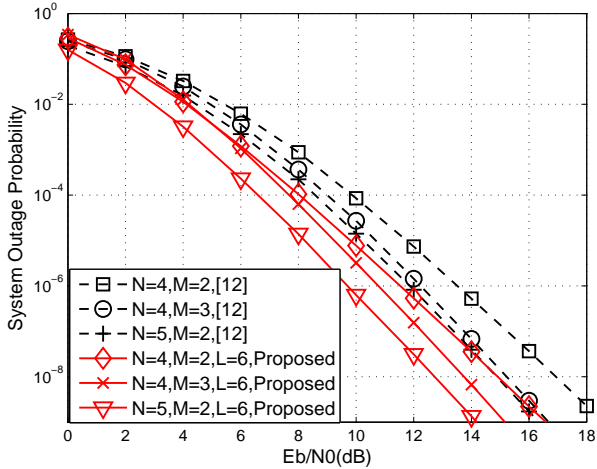}}
\par\end{centering}
\caption{System outage probability performance comparisons between the scheme
of \cite{12} and the proposed US-LCRP under the fixed data rate of
$\mathcal{R}=1bit/s/Hz$. }
\label{fig:DMTComparison_woARA}
\end{figure}

Fig. \ref{fig:DMTComparison_woARA} shows the impact of $M$ and $N$ on the system outage performance
of the proposed US-LCRP under fixed data rate $\mathcal{R}=1bit/s/Hz$.
Observe that the US-LCRP achieves the diversity order of $N+M$, equal
to that of the existing low complexity scheme of \cite{12}. Moreover,
in terms of the system outage probability, the proposed protocol attains
a considerable improvement in comparison with its counterpart in \cite{12}
in terms of the system outage probability. For example, this improvement
is about $1\sim2dB$ when system outage probability is equal to $10^{-4}$
in Fig. \ref{fig:DMTComparison_woARA}.

\begin{figure}[tbp]
\begin{centering}
\textsf{\includegraphics[width=3.5in]{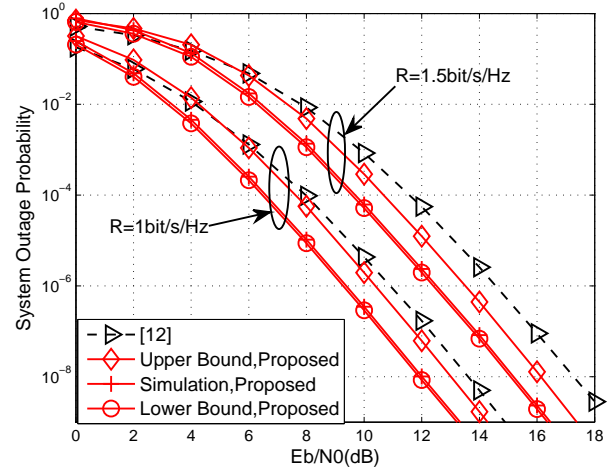}}
\par\end{centering}
\caption{System outage probability performance comparison between the scheme
of \cite{12} and the proposed US-LCRP under the fixed data rate of
$\mathcal{R}=1bit/s/Hz$ and $\mathcal{R}=1.5bit/s/Hz$, respectively
($N=5$, $M=3$).}
\label{fig:ImprovementOfARA}
\end{figure}

Fig. \ref{fig:ImprovementOfARA} shows the impact of data rate $\mathcal{R}$ on the system
outage performance of the existing low complexity scheme \cite{12}
and the proposed US-LCRP. We set $L=6$ in this simulation. The system
outage probabilities of the proposed US-LCRP and of its counterpart
\cite{12} are illustrated in a five-user three-relay network under
fixed data rate $\mathcal{R}=1bit/s/Hz$ and $\mathcal{R}=1.5bit/s/Hz$,
respectively. The upper bound $P_{out}^{'}$ and the lower bound $P_{out}^{''}$
are also illustrated, and the achievable simulation curves of $P_{out}$
reside between them. Thus our analysis of $P_{out}$ is consolidated.
It is shown that when the system outage probability is equal to $10^{-4}$,
the improvement of the proposed US-LCRP over the existing low complexity
scheme of \cite{12} is about $1.5dB$ and $2dB$ for $\mathcal{R}=1bit/s/Hz$
and $\mathcal{R}=1.5bit/s/Hz$ , respectively. We can also observe
that the performance advantage of the proposed scheme over the scheme
of \cite{12} becomes more significant upon increasing $\mathcal{R}$.
This is because the proposed scheme allocates more time resource to
direct transmission (i.e., for the transmission of ``fresh'' data),
and hence the required transmission rate of each link is reduced.
More specifically, in order to achieve an end-to-end data rate of
$\mathcal{R}$ $bit/s/Hz$, the actual data rates of each link in
the proposed scheme and in the scheme of \cite{12} are $\frac{L+M}{L}\mathcal{R}$
$bit/s/Hz$ and $2\mathcal{R}$ $bit/s/Hz$, respectively. Therefore,
the gap of the required data rate in each link between the proposed
scheme and the scheme of \cite{12} grows as $\mathcal{R}$ increases.
As a beneficial result, it is demonstrated that the US-LCRP is more
powerful in supporting high rate transmission than the scheme of \cite{12}.

\begin{figure}[tbp]
\begin{centering}
\textsf{\includegraphics[width=3.5in]{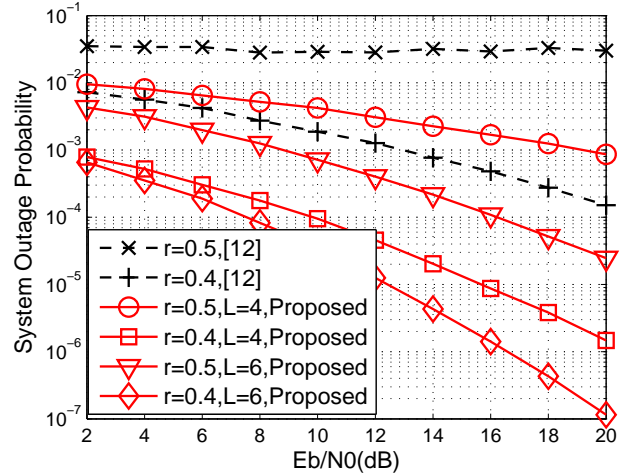}}
\par\end{centering}
\caption{System outage probability performance comparison between the scheme
of \cite{12} and the proposed US-LCRP employed in the asymmetric
network subject to different multiplexing gains ($N=5$, $M=3$).}
\label{fig:ImprovementOfARA-1}
\end{figure}

Fig. \ref{fig:ImprovementOfARA-1} illustrates the system outage probabilities of the proposed
US-LCRP and the scheme of \cite{12} under a range of different values
of multiplexing gain. The desired data rate is determined by the multiplexing
gain and the average SNR of the network as $\mathcal{R}=r\log\left(1+\frac{E_{b}}{N_{0}}\right)$.
It is observed that the US-LCRP achieves higher diversity gain than
the scheme of \cite{12}. For example, if we set $L=6$, the diversity
gains of the US-LCRP are $3.2$ and $2$ when $r=0.4$ and $r=0.5$,
respectively, whereas those of the existing scheme in \cite{12} are
$1.6$ and $0$, respectively. Furthermore, Fig. \ref{fig:ImprovementOfARA-1} shows that the
DMT of the US-LCRP becomes more attractive as $L$ increases. This
result confirms that the proposed scheme has the potential to achieve
better DMT performance than the scheme of \cite{12}. It also implies
that the proposed US-LCRP is the asymptotically optimal in the sense
that the DMT of the US-LCRP becomes arbitrarily close to that of the
ideal MISO scenario when $L$ is sufficiently large.

\begin{figure}[tbp]
\begin{centering}
\textsf{\includegraphics[width=3.5in]{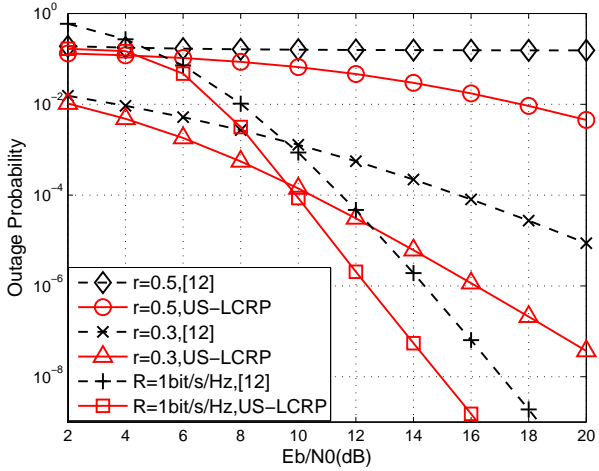}}
\par\end{centering}
\caption{System outage probability performance comparison between the scheme
of {[}12{]} and the proposed US-LCRP subject to different multiplexing
gains ($N=5$, $M=3$, $L=6$).}
\label{fig:Plus}
\end{figure}

Simulation results for the network configured with other parameters
are also presented. For example, in some systems, the relays are almost
in the middle between the source and the destination to assist in
the user's transmission. Hence we assume that the sources are clustered
around $(0,0)$ and the relays are clustered around $(0.5,0.5)$,
whereas the destination remains staying at $(1,1)$. Additionally,
larger path loss exponent, for instance $3.5$, is considered. The
comparison between the proposed US-LCRP and the scheme of {[}12{]}
is depicted in Fig. \ref{fig:Plus}. From Fig. \ref{fig:Plus} we can see that with fixed data
rate, i.e. r=0, the scheme in {[}12{]} and US-LCRP both achieve the
same diversity gain of 8. However, as the multiplexing gain increases,
the diversity gain of the scheme in {[}12{]} decays much faster than
the proposed US-LCRP. Based on the theoretical analysis, the US-LCRP
achieves the diversity gain of $4.4$ and $2$ when $r=0.3$ and $r=0.5$,
respectively, whereas the scheme in {[}12{]} achieves diversity gain
of $3.2$ and $0$, respectively. These results are validated in Fig. \ref{fig:Plus}.

\section{Conclusions}

This paper addresses the problem of joint exploitation of both CD
and MUD under the TP-TRA framework, in which the whole transmission
is composed of the broadcast phase and the relay phase. Based on the
TP-TRA framework, a user-selection based low complexity relay protocol
(US-LCRP) is proposed to achieve both CD and MUD. The DMT performance
of the proposed US-LCRP is analyzed, and simulations are carried out
to corroborate the analysis. Both the theoretical analysis and simulation
results demonstrate that the proposed US-LCRP combines CD and MUD
successfully and achieves a total diversity order of $N+M$ in an
$N$-user $M$-relay network. Furthermore, it provides better DMT
performance and has the potential to approach the optimal DMT when
certain mild conditions are satisfied. Finally, the proposed US-LCRP
needs the CSI of only the direct links in each broadcast TS during
the whole transmission, which results in a simpler implementation.

\appendices{}

\section{Proof of Lemma 1}

Obviously, all the elements in $\mathit{\mathsf{U}}$, which is defined in Section III-A, are mutually
independent random variables (RVs). According to (\ref{eq:SNRsetupdating})
and the definitions in Section III-A, we readily have $\rho_{l}^{\left(m\right)}\mathit{\mathsf{\in U}}$
for all $l$ and $m$ ($1\le l\le L$ and $1\le m\le M$), and $\mathit{\mathsf{V\subseteq U}}$.
From (\ref{eq:UserSelectionRelayPhase}) and (\ref{eq:SNRsetupdating}),
we have $v_{l}=\rho_{l}^{\left(M+1\right)}\geq\rho_{l}^{\left(m+1\right)}=\rho_{l}^{\left(m\right)}>\rho_{\theta_{m}}^{\left(m\right)}$
if $l\neq\theta_{m}$. It means that whenever $x_{l}$ is not selected
in one relay TS, we can find an element $\rho_{\theta_{m}}^{\left(m\right)}\in\mathit{\mathsf{U}}$
that is less than $v_{l}$. Similarly, we obtain $v_{l}=\rho_{l}^{\left(M+1\right)}\geq\rho_{l}^{\left(m+1\right)}>\min\left\{ \rho_{\theta_{m}}^{\left(m\right)},\rho^{\left(m,\mathrm{RP}\right)}\right\} $
if $l=\theta_{m}$. It means that once $x_{l}$ is selected in the
relay phase, an element $\min\left\{ \rho_{\theta_{m}}^{\left(m\right)},\rho^{\left(m,\mathrm{RP}\right)}\right\} \in\mathit{\mathsf{U}}$
is found%
\footnote{We have $\min\left\{ \rho_{\theta_{m}}^{\left(m\right)},\rho^{\left(m,\mathrm{RP}\right)}\right\} \mathsf{\in U}$
because both $\rho_{\theta_{m}}^{\left(m\right)}$ and $\rho^{\left(m,\mathrm{RP}\right)}$
are the elements of $\mathsf{U}$.%
} to be less than $v_{l}$. In summary, after each relay TS, we can
always find an element in $\mathit{\mathsf{U}}$ which is less than
$v_{l}$. Therefore, after the whole relay phase, there are at least
$M$ elements in $\mathit{\mathsf{U}}$ which are less than $v_{l}$,
i.e., $v_{l}^{'}\geq u_{l+M}^{'}$. On the other hand, note that $\mathit{\mathsf{V\subseteq U}}$,
then we have $v_{l}^{'}\leq u_{l+M}^{'}$. Therefore, we obtain the
result of $v_{l}^{'}=u_{l+M}^{'}$, which indicates that $v_{l}^{'}$
is the $\left(M+l\right)$th smallest SNR of all the received signals
during the whole transmission.

\section{Proof of Theorem 1}

It is very difficult to obtain the exact numerical results of (\ref{eq:SymCDF2})
for general networks due to the complicated mathematical structure.
However, in a special network where $\mathrm{F}_{u_{k}}\left(\beta|\theta_{k-L}=l\right)$
remains unchanged with different $l$ (i.e., $\lambda_{S_{i_{l}}R_{m}}$
can be expressed by a constant $\lambda_{SR_{m}}$ for all the $1\leq l\leq L$)
, the expression of $\mathrm{F}_{u{}_{k}}\left(\beta\right)$ can
be derived straightforwardly because $\sum\limits _{l=1}^{L}\Pr\left(\theta_{k-L}=l\right)=1$.
Therefore, $\mathrm{F}_{u_{M+1}^{'}}\left(\beta\right)$ can be obtained
as
\begin{align}
 & \mathrm{F}_{u_{M+1}^{'}}\left(\beta\right)\nonumber \\
= & \sum\limits _{i=M+1}^{L+M}\sum\limits _{{\scriptstyle {\scriptscriptstyle {\textrm{all possible }j_{1},\atop \ldots,j_{L+M}}}}}\Pr\left(u_{j_{1}},\ldots,u_{j_{i}}\leq\beta<u_{j_{i+1}},\ldots,u_{j_{L+M}}\right)\nonumber \\
= & \sum\limits _{i=M+1}^{L+M}\sum\limits _{{\scriptstyle {\scriptscriptstyle {\underset{1\leq j_{i+1}<\ldots<j_{L+M}\leq L+M}{1\leq j_{1}<\ldots<j_{i}\leq L+M}\atop j_{1}\neq\ldots\neq j_{L+M}}}}}\prod\limits _{t=1}^{i}\mathrm{F}_{u{}_{j_{t}}}\left(\beta\right)\nonumber \\
 & \times\prod\limits _{t=i+1}^{L+M}\left(1-\mathrm{F}_{u{}_{j_{t}}}\left(\beta\right)\right).\label{eq:lemma2CDF}
\end{align}
We refer to this special network as the symmetric network. In this
appendix, we show that the DMTs of the two special symmetric networks
serve as the upper bound and the lower bound of that of the original
network, respectively, and the upper bound coincides with the lower
bound. Therefore, we obtain the DMT of the US-LCRP in general cases.

Note that deriving DMT entails the analysis of the asymptotic performance.
Specifically, we can construct some symmetric networks to facilitate
the DMT analysis of the original asymmetric network. For any given
general network, we establish two special $\left(N+M+1\right)$-node
symmetric networks. We assume that the SNRs of the $S_{n}\rightarrow D$,
$S_{n}\rightarrow R_{m}$ and $R_{m}\rightarrow D$ links in the first
symmetric network obey the exponential distributions with parameters
$\lambda_{S_{n}D}^{'}$, $\lambda_{S{}_{n}R_{m}}^{'}$ and $\lambda_{R_{m}D}^{'}$,
respectively, and the SNRs of the $S_{n}\rightarrow D$, $S_{n}\rightarrow R_{m}$
and $R_{m}\rightarrow D$ links in the second symmetric network obey
the exponential distributions with parameters $\lambda_{R_{n}D}^{''}$,
$\lambda_{S{}_{n}R_{m}}^{''}$ and $\lambda_{R_{m}D}^{''}$, respectively.
In the first symmetric network, we set $\lambda_{S_{n}D}^{'}=\lambda_{S_{n}D}$,
$\lambda_{S{}_{n}R_{m}}^{'}=\max\left\{ \lambda_{S{}_{1}R_{m}},\lambda_{S{}_{2}R_{m}},\ldots,\lambda_{S{}_{n}R_{m}}\right\} $
and $\lambda_{R_{m}D}^{'}=\lambda_{R_{m}D}$. Obviously, since the
quality of each link in the first symmetric network is not better
than that of the corresponding link in the original network, it suffers
a higher system outage probability $P_{out}^{'}\left(\mathcal{R}\right)$.
Thus the upper bound of the original network's outage probability
is obtained by calculating the system outage probability of this special
symmetric network using (\ref{eq:lemma2CDF}). Similarly, we can generate
the second symmetric network where the source-to-relay links are guaranteed
to have higher/equal qualities than/to the corresponding links in
the original network, namely we have $\lambda_{S_{n}D}^{''}=\lambda_{S_{n}D}$,
$\lambda_{S{}_{n}R_{m}}^{''}=\min\left\{ \lambda_{S{}_{1}R_{m}},\lambda_{S{}_{2}R_{m}},\ldots,\lambda_{S{}_{n}R_{m}}\right\} $
and $\lambda_{R_{m}D}^{''}=\lambda_{R_{m}D}$. As a result, we can
obtain the lower bound of the original network's outage probability
by calculating the system outage probability $P_{out}^{''}\left(\mathcal{R}\right)$
of the second symmetric network. The mathematical derivation is represented
below.

From \cite{DMT02}, the diversity gain $d$ is computed as
\begin{align*}
d & =-\lim\limits _{\rho\to\infty}\frac{\log P_{out}\left(\mathcal{R}\right)}{\log\rho}=-\lim\limits _{\rho\to\infty}\frac{\log P_{out}\left(r\log\rho\right)}{\log\rho}.
\end{align*}

Since $P_{out}^{'}\left(r\log\rho\right)$ is the upper bound of $P_{out}\left(r\log\rho\right)$,
we have
\begin{equation}
d\geq-\lim\limits _{\rho\to\infty}\frac{\log P_{out}^{'}\left(r\log\rho\right)}{\log\rho}.\label{eq:DMTLB}
\end{equation}

According to (\ref{eq:lemma2CDF}), $P_{out}^{'}\left(r\log\rho\right)$
is computed as
\begin{align}
P_{out}^{'}\left(r\log\rho\right)= & \sum\limits _{l=M+1}^{N+M}\sum\limits _{{\scriptstyle {\scriptscriptstyle {\underset{1\leq j_{l+1}<\ldots<j_{N+M}\leq N+M}{1\leq j_{1}<\ldots<j_{l}\leq N+M}\atop j_{1}\neq j_{2}\neq\ldots\neq j_{N+M}}}}}\prod\limits _{i=1}^{l}\mathrm{F}_{u_{j_{i}}^{\left(UB\right)}}\left(\beta\right)\nonumber \\
 & \times\prod\limits _{i=l+1}^{N+M}\left(1-\mathrm{F}_{u_{j_{i}}^{\left(UB\right)}}\left(\beta\right)\right),\label{eq:UP1}
\end{align}
where we have
\begin{align}
 & \mathrm{F}_{u_{k}^{\left(UB\right)}}\left(\beta\right)\nonumber \\
= & \begin{cases}
\begin{array}{cc}
\prod\limits _{n=1}^{N}\left(1-\exp\left(-\lambda_{S_{n}D}\beta\right)\right), & 1\leq k\leq L\\
\begin{array}{c}
1-\exp\left(-\lambda_{SR_{k-L}}^{'}x-\lambda_{R_{k-L}D}\beta\right)\\
\times K_{1}\left(2\sqrt{\lambda_{SR_{k-L}}^{'}\lambda_{R_{k-N}D}\left(\beta^{2}+\beta\right)}\right)\\
\times2\sqrt{\lambda_{SR_{k-L}}^{'}\lambda_{R_{k-N}D}\left(\beta^{2}+\beta\right)},
\end{array} & {\normalcolor \mathrm{else}}
\end{array} & .\end{cases}\label{eq:UP_X}
\end{align}
Note that $\mathrm{F}_{u_{M+1}^{'}}^{\left(UB\right)}\left(r\log\rho\right)\leq1$,
then we arrive at
\begin{align}
-\lim\limits _{\rho\to\infty}\frac{\log P_{out}^{'}\left(r\log\rho\right)}{\log\rho} & \geq0.\label{eq:DMTLBcase0}
\end{align}
Substituting (\ref{eq:DefSNR}) and (\ref{eq:UP_X}) into (\ref{eq:UP1}),
we have
\begin{align}
 & -\lim\limits _{\rho\to\infty}\frac{\log P_{out}^{'}\left(r\log\rho\right)}{\log\rho}\nonumber \\
= & -\lim\limits _{\rho\to\infty}\frac{\log\left(\rho^{-\left(M+N\right)}\left(\frac{L\beta}{L+M}\right)^{M+N}\right)}{\log\rho}\nonumber \\
= & -\lim\limits _{\rho\to\infty}\frac{\log\left(\rho^{-\left(M+N\right)}\left(\rho^{\frac{L+M}{L}r}\right)^{M+N}\right)}{\log\rho}\nonumber \\
= & -\lim\limits _{\rho\to\infty}\frac{\log\left(\rho^{-\left(M+N\right)\left(1-\frac{L+M}{L}r\right)}\right)}{\log\rho}\nonumber \\
= & \left(M+N\right)\left(1-\frac{L+M}{L}r\right),\label{eq:DMTLBcase1}
\end{align}
if $r<\frac{L}{L+M}$. According to (\ref{eq:DMTLB}) and (\ref{eq:DMTLBcase1}),
we obtain
\begin{align}
d & \geq-\lim\limits _{\rho\to\infty}\frac{\log P_{out}^{'}\left(r\log\rho\right)}{\log\rho}\nonumber \\
 & =\left(M+N\right)\left(1-\frac{L+M}{L}r\right)^{+}.
\end{align}
Similarly, we can get
\begin{align}
d & \leq-\lim\limits _{\rho\to\infty}\frac{\log P_{out}^{''}\left(r\log\left(\rho\right)\right)}{\log\rho}\nonumber \\
 & =\left(M+N\right)\left(1-\frac{L+M}{L}r\right)^{+}.
\end{align}
As a result, the DMT of the proposed US-LCRP is $d=\left(M+N\right)\left(1-\frac{L+M}{L}r\right)^{+}$.

\section*{Acknowledgment}

The authors would like to thank Walaa Hamouda and the anonymous reviewers for their valuable comments.

\bibliographystyle{IEEEtran}


\begin{biography}[{\includegraphics[width=1in,height=1.25in,clip,keepaspectratio]{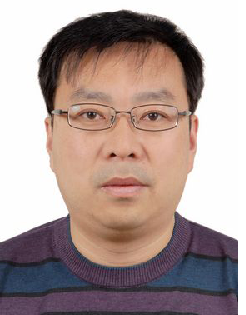}}]{Tiejun Lv}
(M'08-SM'12) received the M.S. and Ph.D. degrees in electronic engineering from the University of Electronic Science and Technology of China (UESTC), Chengdu, China, in 1997 and 2000, respectively. From January 2001 to December 2002, he was a Postdoctoral Fellow with Tsinghua University, Beijing, China. From September 2008 to March 2009, he was a Visiting Professor with the Department of Electrical Engineering, Stanford University, Stanford, CA. He is currently a Full Professor with the School of Information and Communication Engineering, Beijing University of Posts and Telecommunications (BUPT). He is the author of more than 100 published technical papers on the physical layer of wireless mobile communications. His current research interests include signal processing, communications theory and networking.

Dr. Lv is also a Senior Member of the Chinese Electronics Association. He was the recipient of the Program for New Century Excellent Talents in University Award from the Ministry of Education, China, in 2006.
\end{biography}

\begin{biography}[{\includegraphics[width=1in,height=1.25in,clip,keepaspectratio]{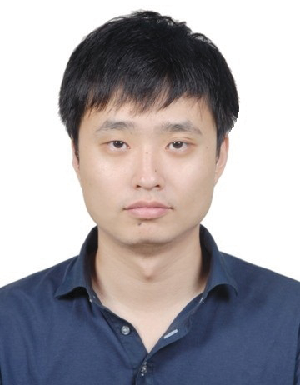}}]{Zhang
Zhang}
received the B.Eng. and Ph.D. degrees from Beijing University of Posts and Telecommunications (BUPT), Beijing, China, in 2007 and 2012, respectively. From May 2009 to June 2012, he also served as a Research Assistant for the Wireless and Mobile Communications Technology R$\&$D Center, Tsinghua University, Beijing, China. He is currently with the Department of Research and Innovation, Alcatel Lucent Shanghai Bell, Shanghai, China.

His current research interests include cooperative communication, network coding, machine type communication and network information theory.
\end{biography}

\begin{IEEEbiography}[{\includegraphics[width=1in,height=1.25in,clip,keepaspectratio]{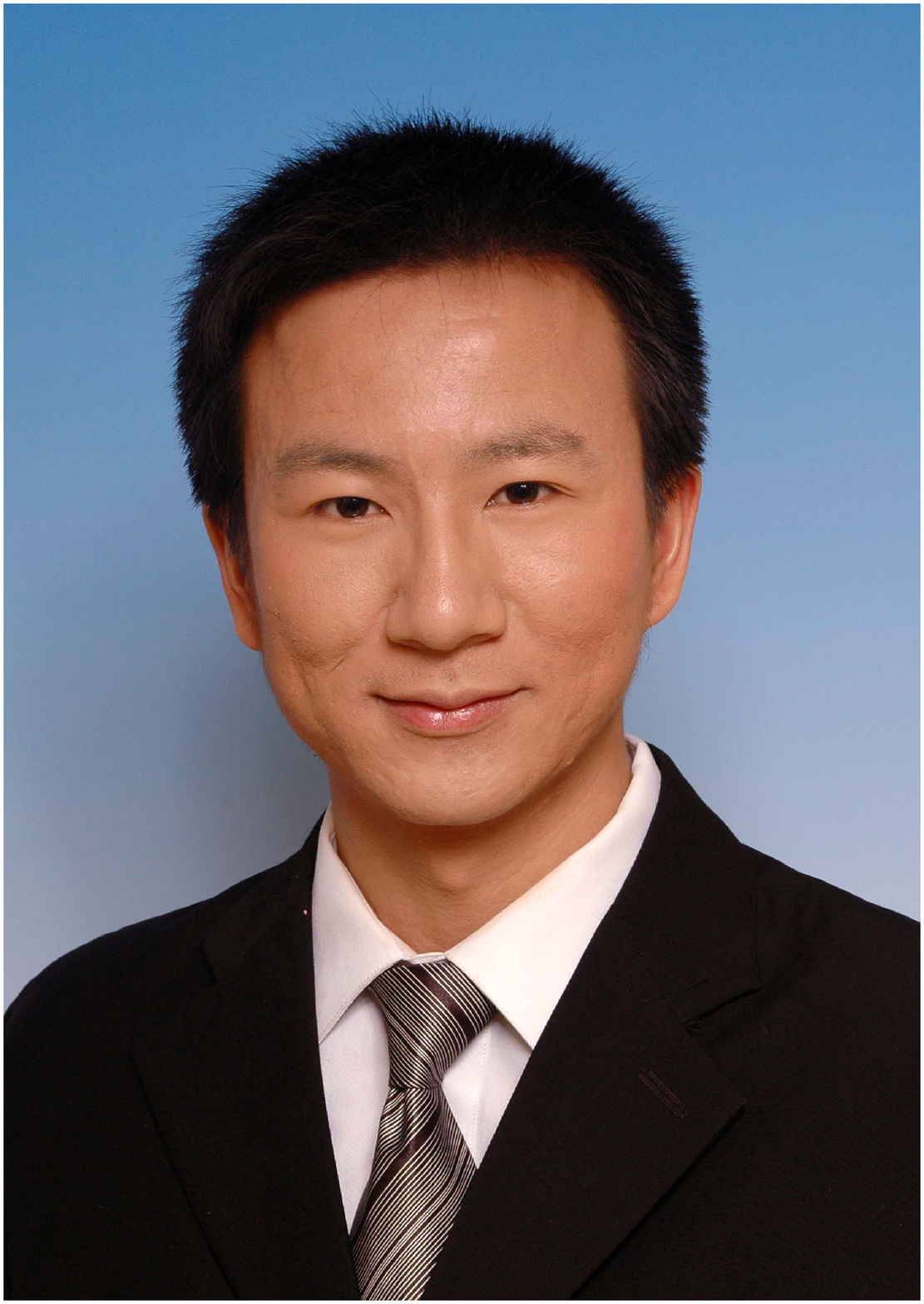}}] {Shaoshi Yang}
(S'09-M'13)   ({https://sites.google.com/site/shaoshiyang/})
received the B.Eng. Degree in information engineering from Beijing University of Posts and Telecommunications, China, in 2006, and Ph.D.
Degree in wireless communications from University of Southampton, U.K., in 2013. He is now working as a Postdoctoral Research Fellow in University of Southampton, U.K.

From November 2008 to February 2009, he was an Intern Research Fellow with the Communications Technology Laboratory, Intel Labs China, Beijing, where he focused on Channel Quality Indicator Channel design for mobile WiMAX (802.16 m).

His research interests include multiple-input--mutliple-output (MIMO) signal processing, multicell joint/distributed signal processing, cooperative communications, green radio, and interference management.
He has published in excess of 25 research papers on IEEE journals and conferences.

Shaoshi is a recipient of the PMC-Sierra Telecommunications Technology Scholarship, and a Junior Member of the Isaac Newton Institute for Mathematical Sciences, Cambridge, UK. He is also a TPC member of the 23rd Annual IEEE International Symposium on Personal, Indoor and Mobile Radio Communications (IEEE PIMRC 2012), of the 48th Annual IEEE International Conference on Communications (IEEE ICC 2013), and of the 2nd IEEE/SAE/ACM/IFAC International Conference on Connected Vehicles and Expo (ICCVE 2013).
\end{IEEEbiography}

\end{document}